\documentclass[reprint,superscriptaddress,aps,prb,floatfix]{revtex4-2}

\usepackage{graphicx}
\usepackage{amsmath}
\usepackage[english]{babel}
\usepackage{units}
\usepackage{mathptmx}
\usepackage[colorlinks=true,urlcolor=blue,linkcolor=blue,citecolor=blue]{hyperref}
\usepackage{xcolor}

\begin{document}

\title{Room-temperature spin-lifetime anisotropy exceeding 60 in bilayer graphene spin valves proximity coupled to WSe$_2$}

\author{Timo Bisswanger}
\affiliation{2nd Institute of Physics and JARA-FIT, RWTH Aachen University, 52074 Aachen, Germany}
\author{Anne Schmidt}
\affiliation{2nd Institute of Physics and JARA-FIT, RWTH Aachen University, 52074 Aachen, Germany}
\author{Frank Volmer}
\affiliation{2nd Institute of Physics and JARA-FIT, RWTH Aachen University, 52074 Aachen, Germany}
\author{Christoph Stampfer}
\affiliation{2nd Institute of Physics and JARA-FIT, RWTH Aachen University, 52074 Aachen, Germany}
\affiliation{Peter Gr\"unberg Institute (PGI-9), Forschungszentrum J\"ulich, 52425 J\"ulich, Germany}
\author{Bernd Beschoten}
\affiliation{2nd Institute of Physics and JARA-FIT, RWTH Aachen University, 52074 Aachen, Germany}
\email{bernd.beschoten@physik.rwth-aachen.de}

\begin{abstract}
A spin lifetime anisotropy between in-plane and out-of-plane spins in bilayer graphene (BLG) can be achieved by spin-orbit proximity coupling of graphene to transition metal dichalcogenides. This coupling reduces the in-plane spin lifetime due to proximity-induced spin scattering, while the out-of-plane spin lifetime remains largely unaffected. We show that at room temperature spin lifetime anisotropy exceeds 60 in a bilayer graphene lateral spin valve proximity coupled to WSe$_2$. The out-of-plane spin lifetime of about \unit[250]{ps} closely matches that of a BLG reference region not in contact with WSe$_2$. In contrast, the estimated in-plane spin lifetime of less than \unit[4]{ps} leads to a complete suppression of the in-plane spin signal measured at the ferromagnetic Co/MgO spin detector. The proximity coupling of  WSe$_2$ to BLG is particularly promising, as it does not compromise the charge carrier mobility within the graphene channel.
\end{abstract}

\maketitle 

\begin{figure*}[bt]
	\includegraphics{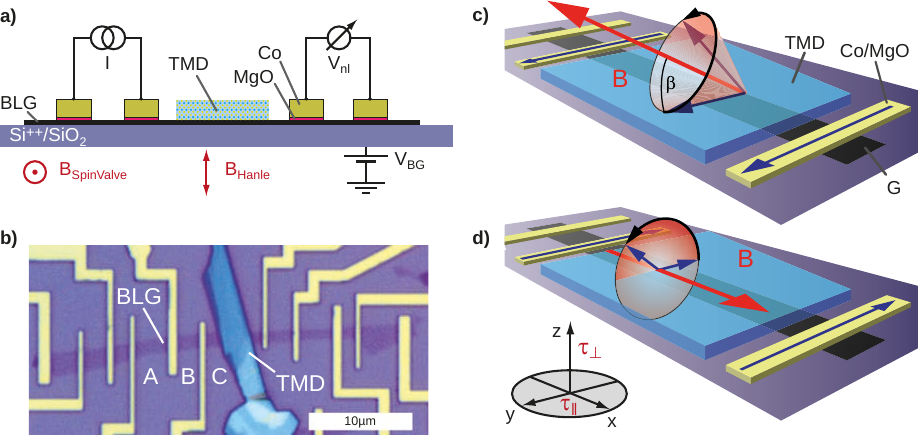}
	\caption{{\bf(a)} Schematic cross-section of the device illustrating the stacking order and the wiring in case of a non-local spin transport measurement. {\bf(b)} Optical image of the device with the reference regions A and B and the WSe$_2$ (TMD) proximity coupled region C. {\bf(c)} In case of the oblique Hanle measurement, the external magnetic field is oriented at an angle $\beta$ between the easy-axis of the ferromagnetic electrodes (blue arrows) ($\beta$\,=\,0$^\circ$) and the out-of-plane field direction of a standard Hanle spin precession measurements ($\beta$\,=\,90$^\circ$). {\bf(d)} In the \textit{x}-Hanle configuration the external magnetic field points in the \textit{x}-direction, which is defined as the direction perpendicular to the ferromagnetic contacts and within the plane of the graphene channel.}
	\label{fig:fig1}
\end{figure*}

\section{Introduction}

A promising method to manipulate the spin transport in graphene-based spin-valve devices is to couple the graphene transport channel to a material with high spin-orbit coupling~\cite{Gmitra-PRB-2015,RocheAkermann-2DMater-2015,Zollner2025May}. Such materials include transition metal dichalcogenides (TMDs) like WSe$_2$, WS$_2$, MoSe$_2$, MoS$_2$ or their alloys~\cite{Gmitra-PRL-2017,WangMorpurgo-PRX-2016,PhysRevB.106.125417}, but also materials like topological insulators or two-dimensional magnetic insulators, where the latter includes composites such as Cr$_2$Ge$_2$Te$_2$ or MnPSe$_3$ that can induce ferro- or antiferromagnetic proximity exchange coupling in graphene~\cite{2DMaterials.12.013004,2DMaterials.7.015026,PhysRevB.107.195144}. Due to their strong SOC, TMDs that are proximity-coupled to graphene imprint an effective out-of-plane spin orbit field on the graphene layer, which enhances spin scattering for spins oriented in the graphene plane while leaving out-of-plane spins largely unaffected~\cite{Cummings-PRL-2017, Offidani-PRB-2018,Gmitra-PRB-2015,Gmitra-PRL-2017,WangMorpurgo-PRX-2016,2DMaterials.12.013004}. This yields a pronounced spin lifetime anisotropy, defined as $\xi=\tau_\perp/\tau_\parallel$, between the out-of-plane ($\tau_\perp$) and in-plane ($\tau_\parallel$) spin relaxation times \cite{APLMaterials.7.120701}.

In lateral spin-valve experiments on TMD/single-layer graphene (SLG) heterostructures utilizing MoS$_2$, MoSe$_2$, WS$_2$, and PdSe$_2$ spin-lifetime anisotropies of about $\xi\approx10$ have been reported so far \cite{Benitez-NatPhys-2018,Ghiasi-NanoLett-2017, PhysRevMaterials.7.044005,NatureMaterials.24.876}, whereas the theoretical expectation is at least one order of magnitude higher \cite{Cummings-PRL-2017}. Furtheremore, the anisotropy in these studies was not sufficiently high enough to completely suppress the in-plane spin signal from reaching the detector electrode \cite{Benitez-NatPhys-2018,Ghiasi-NanoLett-2017, PhysRevMaterials.7.044005,NatureMaterials.24.876}. Even though WSe$_2$ has already been used in lateral spin valves, no anisotropy was determined or quantified that way \cite{Ingla-Aynes-PRL-2021, Herling-APLMater-2020}, but in weak antilocalization measurements on WSe$_2$/SLG/hBN-heterostructures at temperatures below \unit[2]{K} a lower bound of $\xi=20$ has been found \cite{Zihlmann-PRB-2018}.

Furthermore, using bilayer graphene (BLG) instead of single layer graphene offers additional opportunities for graphene-based spintronic devices that are proximity coupled to TMDs. Electrical transport measurements in BLG devices that are in contact with TMDs on either one or both sides reveal that the proximity-induced SOC is layer-selective, primarily impacting the graphene layer in direct contact with the TMD~\cite{Nature.571.85,2DMaterials.12.035009,Icking2025May}. Combined with the fact that the band structure of BLG can be tuned with out-of-plane electric fields to either open a band-gap~\cite{Eike-gap-paper} or to confine all charge carriers to only one of the two layers~\cite{Khoo-NanoLett-2017}, enables device concepts such as spin-orbit valves or spin transistors~\cite{Khoo-NanoLett-2017,Gmitra-PRL-2017}.
In this respect, a BLG spin-valve that was fully supported by a large WS$_2$ flake has already shown a spin-lifetime anisotropy of about $\xi=40\dots70$ albeit at low temperatures of~\unit[4]{K}~\cite{Omar-PRB-2019}. However, much higher values up to $\xi=10\,000$ were theoretically predicted in BLG proximity-coupled to WSe$_2$~\cite{Gmitra-PRL-2017}. At the same time, WSe$_2$ is a substrate that leaves the charge carrier mobility largely unaffected in its graphene-based heterostructures~\cite{Banszerus-2DMater-2017, Banszerus-arXiv-2019}. Therefore, WSe$_2$ is expected to be particularly suitable for introducing high spin lifetime anisotropy into graphene-based spin valves through proximity coupling, while retaining its charge transport properties.

Here we report a room temperature spin-lifetime anisotropy with a lower bound of 60 obtained in a lateral spin-valve device consisting of bilayer graphene coupled to WSe$_2$. The high anisotropy leads to a complete suppression of the in-plane spin signal measured at the ferromagnetic detector electrode in both in-plane Hanle and spin-valve measurements. Instead, the spin signal could be recovered in oblique Hanle measurements and \textit{x}-Hanle measurements involving the out-of-plane spin component \cite{Raes-NatComm-2016, Raes-PRB-2017, Zhu-Probing-PRB-2018, Xu-aniso-PRL-2018}. Because of the full suppression of the in-plane component, the respective lifetime was estimated by numerical simulations of the angle-dependent, anisotropic spin precession data to be less than \unit[4]{ps}.

\section{Results and Discussion}

The device discussed here is built from exfoliated bilayer graphene on a Si$^{++}$/SiO$_2$ (285~nm) substrate, where one part of the BLG transport channel is proximity coupled to a multilayer WSe$_2$ flake with a thickness between 4-7~nm determined by optical contrast measurements in the heterostructure region (see Figs.\,\ref{fig:fig1}a and \ref{fig:fig1}b). The WSe$_2$ flake was placed across the BLG with a dry transfer technique using a PDMS-droplet \cite{Bisswanger-NanoLett-2022} and is located in sample region C in Fig.\,\ref{fig:fig1}b, further referred to as the "WSe$_2$ region". Two uncovered regions labeled A and B in Fig.\,\ref{fig:fig1}b serve as reference regions to characterize spin transport. The spin-sensitive electrodes were fabricated from \unit[3]{nm} magnesium oxide (MgO), serving as spin injection/detection barriers, and \unit[35]{nm} cobalt (Co) \cite{Volmer-PRB-2013,Volmer-PRB-2014}, both deposited under ultra-high vacuum conditions \cite{Volmer-HowSolveProblems-2021}. The contact-resistance-area products of the fabricated electrodes were in the range of 2.4-7.2~k$\Omega\mu$m$^2$. The width of the BLG flake in the investigated regions is in the range of 1.1-1.5~$\mu$m and the contacts have a width of \unit[450-800]{nm} in the \textit{x}-direction (see Fig.\,\ref{fig:fig1}d for coordinate system). For such a contact geometry only a single magnetic domain is formed with its easy-axis along the \textit{y}-direction \cite{Leven-PRB-2005}. A magnetic field in either the \textit{x}-direction (in-plane perpendicular to the contacts) or in the \textit{z}-direction (out-of-plane) slowly rotates the magnetization from the easy-axis with increasing field strength \cite{Brands-JAP-2005}, which we accounted for in our data analysis by a Stoner-Wohlfarth model \cite{Stoner-Wohlfarth-1948,Leutenantsmeyer-PRL-2018,Ghiasi-NanoLett-2017}. All measurements were performed at room temperature. For the non-local spin transport measurements a low-frequency lock-in technique was used with a frequency of $f= \unit[21]{Hz}$ and an AC current of $I_{\mathrm{AC}}$\,=\,\unit[10]{$\mu$A} added on a DC current of $I_{\mathrm{DC}}$\,=\,\unit[25]{$\mu$A}~ \cite{Volmer-PRAppl-2022}.

The WSe$_2$ region exhibits a charge carrier mobility of \unit[5\,800]{cm$^2$/(Vs)}, as determined from gate-dependent conductance measurements \cite{Bisswanger-NanoLett-2022}. The mobility is comparable to mobility values of \unit[5\,200]{cm$^2$/(Vs)} and \unit[6\,800]{cm$^2$/(Vs)} for the reference regions A and B, respectively. We therefore conclude that the presence of WSe$_2$ in these heterostructures on SiO$_2$ the WSe$_2$ does not adversely affect the charge carrier mobility in BLG. This observation is consistent with previous studies on high-quality, fully encapsulated graphene devices, which demonstrated that WSe$_2$ substrates allows for charge carrier mobilities comparable to, or even exceeding, those achieved with hexagonal boron nitride (hBN) encapsulation \cite{Banszerus-arXiv-2019, Ouaj2024Dec}.

\subsection{Spin transport in the reference regions}
\begin{figure}[tb]
    \includegraphics{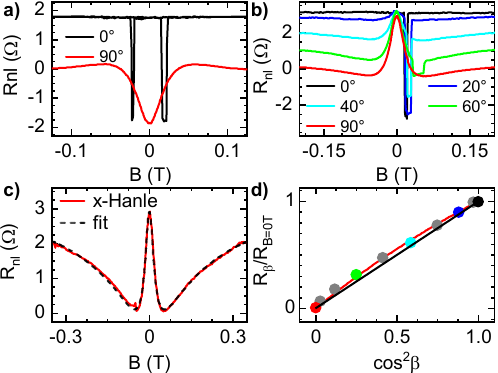}
    \caption{
    Results for BLG reference regions. 
    {\bf (a)} \textit{z}-Hanle (90$^\circ$) and spin-valve measurements (0$^\circ$) of reference region B. {\bf (b)} oblique Hanle in reference region A at selected angles. {\bf (c)} \textit{x}-Hanle measurement with fit (black dotted curve), and {\bf (d)} the evaluation of the anisotropy from the oblique Hanle measurements for reference region A. The colored dots relate to the curves shown in (b), the black straight line corresponds to $\xi=1$, and the red line is the fit to the data.}
    \label{fig:fig2}
\end{figure}

We first study the spin transport properties of the reference regions by performing spin-valve and Hanle measurements at room temperature (RT) at zero gate voltage near the charge neutrality point in a non-local configuration (Fig.\,\ref{fig:fig1}a). For the spin-valve measurements, the magnetic field is changed in its magnitude along the \textit{y}-direction (see Fig.\,\ref{fig:fig1}c), i.e., at angle $\beta=0^\circ$ between the \textit{y} and \textit{z}-direction. Fig.\,\ref{fig:fig2}a shows the measurement of region A. The spin-valve measurement (black curve) shows typical switching with a non-local resistance $\Delta R_\text{nl}$ of about \unit[2.5]{$\Omega$} (\unit[3.6]{$\Omega$} for region B). From the spin precession measurements in Hanle configuration with a magnetic field perpendicular (i.e., $\beta=90^\circ$) to the sample plane (red curve in Fig.\,\ref{fig:fig2}a), the spin lifetime $\tau_s$ and spin diffusion length $\lambda_s$ were determined with the steady-state solution of the Bloch-Torrey equation \cite{JohnsonSilsbee-PRB-1988,FabianZutic-Semiconductor-2007}. To better distinguish the different measurement configurations, we will further refer to this as \textit{z}-Hanle. The corresponding spin lifetime is \unit[360]{ps} (\unit[280]{ps} for region B) with a spin diffusion length of $\lambda_s$=\unit[2.8]{$\mu$m} (\unit[1.7]{$\mu$m} for region B).

In the \textit{z}-Hanle measurement, the spin precession takes place entirely in the sample plane (\textit{x-y}-plane). Therefore, to measure the spin-lifetime anisotropy, the spins must acquire an out-of-plane component during their precession. This can be achieved with two different measurement schemes: (1) In oblique Hanle measurements (Fig.\,\ref{fig:fig1}c) the magnetic field is applied at an angle $\beta$ between $\beta=0^\circ$ and 90$^\circ$, so that an angle-dependent out-of-plane component is obtained due to the tilted precession of the spins. (2) For \textit{x}-Hanle, the magnetic field is applied in the \textit{x}-direction, i.e., in-plane but perpendicular to the contacts, which is illustrated in Fig.\,\ref{fig:fig1}d. During precession, the spins gain an out-of-plane component as they now precess in the \textit{y-z} plane. 

For reference region A, we discuss first the oblique Hanle measurements. A selection of the corresponding curves recorded at different angles $\beta$ is given in Fig.\,\ref{fig:fig2}b. The curves appear as a combination of spin-valve (0$^\circ$) and z-Hanle measurements (90$^\circ$). The spin-lifetime anisotropy was determined following the evaluation proposed by Refs.\,\cite{Raes-PRB-2017} and \cite{Leutenantsmeyer-PRL-2018}: At higher magnetic fields (\unit[150]{mT}), the spins are assumed to be fully dephased, leading to a saturation value that depends on angle and anisotropy. The dependence on anisotropy as derived in \cite{Raes-NatComm-2016,Raes-PRB-2017} is as follows:
\begin{equation}
  \label{eqn:cos2beta}
  \frac{R_{\mathrm{nl}\beta}}{R_{\mathrm{nl}0}}=\chi^{\frac{1}{2}}\exp{\left[\frac{-L}{\lambda_\parallel}\left(\chi^{-\frac{1}{2}}-1\right)\right]}\cos^2{\beta},
\end{equation}
where 
\begin{equation}
  \label{eqn:xitilde}
  \chi
  =\left(
  \cos^2{\beta}
  +\xi^{-1}\sin^2{\beta}\right)^{-1},
\end{equation}
where $L$ is the channel length, $\lambda_\parallel=\sqrt{\tau_\parallel D_s}$ represents the in-plane spin diffusion length, $R_{\mathrm{nl}\beta}$ is the non-local resistance in the saturation regime measured at angle $\beta$, and $R_{\mathrm{nl}0}$ is the value at $B=0$. The spin lifetime anisotropy $\xi$ becomes the only fit parameter when $\tau_\parallel$ and $D_s$ are determined independently from the \textit{z}-Hanle measurement. 

The saturation resistance of the reference region $R_{\mathrm{nl}\beta}$ is normalized with $R_{\mathrm{nl}0}$ and plotted in Fig.\,\ref{fig:fig2}d against $\cos^2(\beta)$. In an isotropic case ($\tau_\perp=\tau_\parallel$), a linear relationship versus $\cos^2\beta$ is expected as indicated by the straight line in Fig.\,\ref{fig:fig2}d \cite{Benitez-NatPhys-2018}. For anisotropies $\xi > 1$ ($< 1$), the curve would be above (below) the $\xi=1$ curve (black line). The fit to our data (red line) gives an anisotropy of $\xi=1.19$. According to Ref. \cite{Raes-NatComm-2016} a value slightly above 1 can be explained by small out-of-plane spin-orbit fields that are caused by ripples or flexural distortions. However, in our sample we also observe that the oblique Hanle curves do not completely saturate even at higher magnetic fields as can be seen in the non-vanishing slope above \unit[150]{mT} in all curves with $\beta\neq 0$ in Fig.\,\ref{fig:fig2}b. This slope can be explained by a Hall signal, that is caused by a spatially inhomogeneous injection of charge carriers through pinholes in the MgO barrier \cite{Volmer-2DMater-2015}. Since this  Hall signal depends on the magnetic field direction (see \cite{Volmer-2DMater-2015}), it also varies with $\beta$ and therefore influences the previously described  method for evaluating the spin lifetime anisotropy.

To investigate this possible source of error with regard to the extracted spin lifetime anisotropy further, we performed \textit{x}-Hanle measurements (Fig.\,\ref{fig:fig2}c), in which the in-plane direction of the magnetic field does not result in any Hall voltages. In the isotropic case, the shape of the curve is expected to be identical to the \textit{z}-Hanle curve, since the projection of the precessing spin ensemble on the \textit{y}-direction (direction of the magnetization of the contacts) is identical regardless if the spins precess in the \textit{z-y-} or \textit{x-y}-plane. However, due to the shape anisotropy of the contacts, the magnetization of the contacts rotates significantly more when applying the magnetic field along the \textit{x}-direction than along the \textit{z}-direction \cite{Brands-JAP-2005}. The more the magnetization rotates towards the magnetic field direction, the larger the overall contribution of the injected spins (which are injected parallel to the magnetization) that is parallel to the external field. This non-precessing contribution of the injected spins results in an additional background signal, that increases the more the magnetization is rotated away from the \textit{y}-direction. This background is seen in a Fig.\,\ref{fig:fig2}c as a steadily growing signal away from $B=0$ in both directions, while $R_\mathrm{nl}$ above $|B|>\unit[150]{mT}$ in the \textit{z}-Hanle configuration approaches $R_\mathrm{nl}=0$ (see Fig.\,\ref{fig:fig2}a).

We implemented a Stoner-Wohlfarth model \cite{Stoner-Wohlfarth-1948,Leutenantsmeyer-PRL-2018,Ghiasi-NanoLett-2017,Ringer-PRB-2018} in the steady-state solution of the Bloch-Torrey equation \cite{FabianZutic-Semiconductor-2007} to account for the rotation of the magnetization of the contacts in the \textit{x}-Hanle measurement scheme. With the adapted fit (see dashed black line in Fig.\,\ref{fig:fig2}c), we find a spin-lifetime of \unit[350]{ps} in reference region A, which is within the fit error identical to the lifetime extracted from the \textit{z}-Hanle measurement (\unit[360]{ps}). This implies an anisotropy factor of 1, as expected from graphene without any proximity coupling \cite{Leutenantsmeyer-PRL-2018}, in contrast to the value of 1.19 extracted from the oblique Hanle method. This indicates that charge-induced Hall signals must be considered if precise anisotropy values are to be extracted from oblique Hanle measurements.

\begin{figure}[th]
	\includegraphics{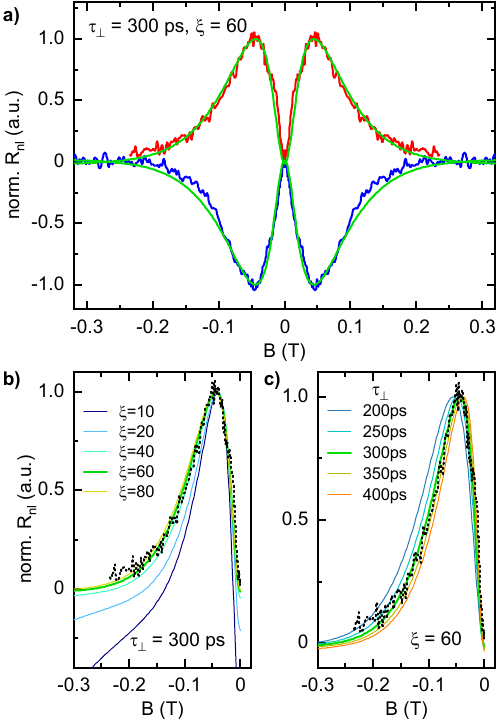}
	\caption{\textit{x}-Hanle measurement and its numerical simulations in the WSe$_2$ region.
    {\bf (a)} \textit{x}-Hanle measurement and best simulation result (green lines) with $\tau_\perp$=\unit[300]{ps} and $\xi=60$. {\bf (b)} Variation of the spin-lifetime anisotropy $\xi$ for a constant $\tau_\perp$=\unit[300]{ps}.
    {\bf (c)} Variation of $\tau_\perp$ for a constant $\xi=60$. The spin diffusion coefficient is set to 0.025~m$^2$/s for the simulations.} 
	\label{fig:fig3}
\end{figure}

\subsection{WSe$_2$ proximity-coupled region}

\begin{figure*}[th]
	\includegraphics{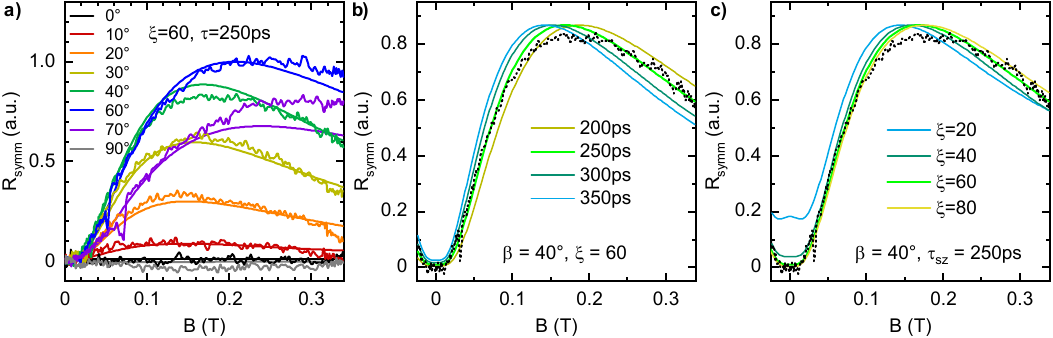}
	\caption{Symmetrized and normalized oblique Hanle measurements of the WSe$_2$-region and simulation curves. {\bf (a)} Measurement set at different angles $\beta$ with corresponding simulations using the best match parameters from (b) and (c), i.e.\ $\tau_\perp$=\unit[250]{ps} and $\xi=60$. {\bf (b)} Measurement at $\beta=40^\circ$ and simulated curves for which the spin-lifetime anisotropy $\xi=60$ was kept constant while varying the out-of-plane spin lifetime $\tau_{\perp}$. {\bf (c)} Similarly to (b) $\tau_{\perp}$=\unit[250]{ps} is kept constant while varying $\xi$. The spin diffusion coefficient is set to 0.025~m$^2$/s for the simulations.}
	\label{fig:fig4}
\end{figure*}

In contrast to the uncovered BLG regions A and B, we could not detect any spin signal in the WSe$_2$ region in either spin-valve configuration ($\beta=0^\circ$) or \textit{z}-Hanle configuration ($\beta=90^\circ$) (see black and gray lines in Fig.\,\ref{fig:fig4}a). Spins injected in-plane obviously do not reach the detector as long as their orientation remains in-plane due to an extremely short in-plane spin lifetime. Since higher out-of-plane spin lifetimes are expected within the WSe$_2$ proximity-coupled region, we next performed oblique Hanle and \textit{x}-Hanle measurements in the electron conduction regime at a charge carrier density of $n=\unit[2\times 10^{12}]{cm^{-2}}$.

Fig.\,\ref{fig:fig3}a shows the \textit{x}-Hanle measurements for both parallel and antiparallel orientation of the injection and detection electrode's magnetization. To limit the impact of charge-induced background signals, the spin signal for each orientation was symmetrized and normalized according to $R_\mathrm{symm}:=(|R_\mathrm{nl}(-B)|+|R_\mathrm{nl}(B)|)/2-R_\mathrm{nl}(B=0)$. As expected from the \textit{z}-Hanle measurements, the spin signal is zero at $B=\unit[0]{T}$. Instead, for increasing magnetic field strength the spin signal first increases due to an increased out-of-plane contribution, before it decreases again due to magnetic field-induced dephasing of the spin ensemble (see also Ref.\,\cite{Ghiasi-NanoLett-2017}). Unfortunately, fitting these curves with the anisotropic solution of the Bloch-Torrey equation proposed in Ref.\,\cite{Xu-aniso-PRL-2018} is not possible, as this fitting procedure requires a pre-determined in-plane spin lifetime and spin diffusion coefficient extracted from \textit{z}-Hanle measurements, which is not possible in our device as the in-plane spin transport is fully suppressed.

However, a numerical simulation of the anisotropic spin-precession data is possible, using a three-dimensional approach of the Bloch-Torrey equation in COMSOL Multiphysics\textregistered\ software, adapted from the model presented in Ref.\,\cite{Ringer-PRB-2018}. Assuming that charge and spin diffusion coefficients are identical for each region, we use the previously determined spin diffusion coefficient of the reference region in our simulations due to similar mobilities between the regions. The best simulation of the \textit{z}-Hanle measurement was achieved by assuming an out-of-plane spin lifetime of $\tau_\perp=\unit[300]{ps}$ with a spin lifetime anisotropy of $\xi=60$ (green curve in Fig.\,\ref{fig:fig3}a). For comparison, Fig.\,\ref{fig:fig3}b shows simulations with a constant out-of-plane spin lifetime of \unit[300]{ps} but with different spin lifetime anisotropies $\xi=10\dots80$. We find that the simulation approximates the shape of the curve well for values of $\xi\geq 60$ without being able to give an upper limit. Similarly, Fig.\,\ref{fig:fig3}c depicts the simulation results for the estimated anisotropy of $\xi=60$ for varying out-of-plane spin lifetimes.

In a next step, we conducted oblique Hanle measurements in the WSe$_2$ region as shown in Fig.\,\ref{fig:fig4}a. The respective measurements in spin-valve and \textit{z}-Hanle configuration, corresponding to $\beta=0^\circ$ and 90$^\circ$, show no features above the noise level, as mentioned earlier. Instead, a spin signal appears for intermediate angles. Typical evaluation methods for this kind of data set as proposed in Refs.\,\cite{Raes-NatComm-2016,Raes-PRB-2017,Zhu-Modeling-PRB-2018, Xu-aniso-PRL-2018} are again not applicable in our case, as they require a residual spin signal for $\beta=0^\circ$ and 90$^\circ$ to obtain reference values. We therefore again apply numerical simulations to find a best set for $\tau_\perp$ and $\xi$ that can describe the entire set of curves. 

Starting with the best match values from the \textit{x}-Hanle simulations, we examine the simulation results at an angle of $\beta=40^\circ$ for different out-of-plane spin lifetimes for a fixed value of the anisotropy of $\xi=60$ in Fig.\,\ref{fig:fig4}b. Only a lifetime of $\tau_\perp=$\unit[250]{ps} (green curve) follows the experimental data (black points) reasonably well over the whole magnetic field range. Shorter lifetimes underestimate the data for smaller fields and overestimate it for higher fields. The exact opposite (e.g.\;overestimation for small fields and underestimation for higher fields) is seen in case of higher spin lifetimes.

Fig.\,\ref{fig:fig4}c shows the corresponding simulation results for different spin lifetime anisotropies with a fixed spin lifetime of $\tau_\perp$=\unit[250]{ps}. The variation of $\xi$ has its most pronounced effect around $B=\unit[0]{T}$. For smaller values of $\xi$, a residual in-plane spin signal in the form of a small Hanle peak appears at $B=\unit[0]{T}$, as measured in previous studies \cite{Benitez-NatPhys-2018, Leutenantsmeyer-PRL-2018,Omar-PRB-2019}. As was the case with the \textit{x}-Hanle simulations, we find that the simulation approximates the shape of the curve well for values of $\xi\geq 60$ without being able to give an upper limit. The best-match values obtained this way ($\xi=60$ and $\tau_\perp$=\unit[250]{ps}), also quite well simulate the oblique data set for every other angle $\beta$ (see solid lines in Fig.\,\ref{fig:fig4}a). Smaller deviations between experimental data and simulations can be explained by signals caused by the Hall effect \cite{Volmer-2DMater-2015,Volmer-PRAppl-2022}, which was not considered in our model.

\subsection{Underestimation of the spin lifetime anisotropy}

Although the extracted spin lifetime anisotropy of $\xi\geq 60$ is one of the highest values reported in literature so far, it remains well below the theoretical prediction of up to $\xi=10\,000$ in case of BLG proximity-coupled to WSe$_2$~\cite{Gmitra-PRL-2017}. The most likely explanation for this is that we have neglected the impact of contacts and substrates on the measured spin lifetimes. It has been demonstrated that using the steady-state solution of the Bloch-Torrey equation~\cite{JohnsonSilsbee-PRB-1988,FabianZutic-Semiconductor-2007} without considering contact- or substrate-induced spin relaxation processes leads to an underestimation of the extracted spin lifetime~\cite{Maassen-PRB-2012,Stecklein-PRAppl-2016,Volmer-PRB-2014,Droegeler-PSSB-2017,Sosenko-PRB-2014,Zhu-Modeling-PRB-2018,Amamou-ContactInducedSpin-2016,PhysRevB.105.115122,VanTuan-SciRep-2016,Ertler-PRB-2009}.

The measured total spin relaxation rate (i.e.~the inverse of the measured spin lifetime) is the sum of contributions from spin relaxation rates due to the contacts, the SiO$_2$ substrate and the proximity-induced out-of-plane spin orbit field caused by the TMD. Assuming that only the latter exhibits an anisotropy between out-of-plane and in-plane spin lifetimes, whereas the spin relaxation rates due to the SiO$_2$ substrate and the MgO/Co contacts are the same for both spin orientations, we can make the following reasoning: In case of the in-plane spin lifetime, the complete suppression of the in-plane spin component implies that the proximity-induced out-of-plane spin orbit field completely dominates over other relaxation channels. The opposite is true for the out-of-plane spin lifetime, at it is very similar to the spin lifetime in the reference region, implying that substrate- and contact-induced relaxation rates dominate here.

At the same time, it was shown that spin lifetimes of up to 12~ns can be reached in graphene-based spin valve devices~\cite{Droegeler-NanoLett-2016}. This implies that the determined out-of-plane spin lifetime of 250–300~ps could, in principle, be enhanced by at least a factor of 40, if substrate- and contact-induced relaxation processes could be diminished. Under such conditions, the corresponding spin lifetime anisotropy could potentially increase to $\xi=60\times40=2\,400$, bringing it much closer to the theoretically expected value of $\xi=10\,000$.

Furthermore, we note that the twist angle between the TMD and BLG might also have an impact on the anisotropy. Prior studies have demonstrated that the strength of the SO proximity effect depends on the twist angle \cite{Li-Koshino-PRB-2019,Pezo-2DMater-2021,Peterfalvi-PRR-2022,Lee-deSousa-PRB-2022, Zollner2023Dec,PhysRevB.107.195144,NatureMaterials.23.1502,PhysRevB.109.L241403}. In our sample, we estimate the twist angle between the flakes to be approximately 18$^\circ$, based on microscope images and the assumption that the flakes have ripped along the same crystallographic axes during exfoliation. This estimated angle falls in the regime of strong Valley-Zeeman coupling \cite{Li-Koshino-PRB-2019,Lee-deSousa-PRB-2022} which favors large spin lifetime anisotropies. However, further studies are required to determine the dependence of twist angle on spin lifetime anisotropy.

\subsection{Conclusions}

In summary, we have demonstrated a large room temperature spin lifetime anisotropy of $\xi \geq 60$ in bilayer graphene spin valves proximity-coupled to WSe$_2$. Through a combination of Hanle and oblique spin precession measurements and supported by numerical simulations, we determined that the out-of-plane spin lifetime in the BLG/WSe$_2$ region is approximately 250–300~ps, whereas the in-plane spin lifetime is suppressed below 4 ps. The latter effectively eliminates transport of in-plane spins, allowing only the out-of-plane oriented spins to propagate in the direction of the spin detector. The achieved spin lifetime anisotropy of 60 is the highest at 300~K reported so far for graphene-based heterostructures. At the same time, the determined anisotropy factor represents a conservative lower bound of what can be achieved in a BLG/WSe$_2$ heterostructure, as substrate- and contact-induced spin relaxation processes have likely limited the measured out-of-plane spin lifetime. We attribute the large spin lifetime anisotropy to the use of BLG, which is predicted to exhibit significantly higher anisotropy compared to SLG \cite{Gmitra-PRL-2017, Omar-PRB-2019}.

\begin{acknowledgments}
This project has received funding from the European Union's Horizon 2020 research and innovation programme under grant agreement No. 881603 (Graphene Flagship), the Deutsche Forschungsgemeinschaft (DFG, German Research Foundation) under Germany's Excellence Strategy - Cluster of Excellence Matter and Light for Quantum Computing (ML4Q) EXC 2004/1 - 390534769 and by the Helmholtz Nano Facility~\cite{HNF} at the Forschungszentrum J\"ulich.
\end{acknowledgments}

{\bf Author contributions:}
T.\,Bisswanger and A.\,Schmidt contributed equally to this work.

{\bf Notes:} The data that support the findings of this study are openly available at https://doi.org/10.5281/zenodo.15609970.

\end{document}